\RequirePackage[2020-02-02]{latexrelease}

\documentclass[twocolumn,preprintnumbers]{revtex4}%
\usepackage{amssymb}
\usepackage{amsmath}
\usepackage{graphicx}
\usepackage{dcolumn}
\usepackage{bm}
\usepackage{xcolor}
\usepackage{ifthen}
\usepackage[normalem]{ulem}
\usepackage{amsfonts}%
\RequirePackage[2020-02-02]{latexrelease}
\UseRawInputEncoding
\providecommand{\U}[1]{\protect\rule{.1in}{.1in}}
\providecommand{\U}[1]{\protect\rule{.1in}{.1in}}
\def\showal{1}
\newcommand{\al}[1]{\ifthenelse{\showal=1}{\textcolor{orange}{[[#1]]}}{}}

\newcommand{\eb}[1]{\ifthenelse{\showal=1}{\textcolor{cyan}{[[#1]]}}{}}
\begin{document}
\title{Disentanglement by deranking and by suppression of correlation}
\author{Eyal Buks}
\email{eyal@ee.technion.ac.il}
\affiliation{Andrew and Erna Viterbi Department of Electrical Engineering, Technion, Haifa
32000, Israel}
\date{\today }

\begin{abstract}
The spontaneous disentanglement hypothesis is motivated by some outstanding
issues in standard quantum mechanics, including the problem of quantum
measurement. The current study compares between some possible methods that can
be used to implement the hypothesis. Disentanglement is formulated using a
nonlinear operator, which can be used to modify both the Schr\"{o}dinger
equation for the quantum state vector, and the master equation for the density
operator. Two types of nonlinear disentanglement operators are explored. The
first one gives rise to matrix deranking, and the second one to correlation
suppression. Both types are demonstrated using a two spin system that is
driven close to the Hartmann--Hahn double resonance. It is shown that limit
cycle steady state solutions, which are excluded by standard quantum
mechanics, become possible in the presence of disentanglement.

\end{abstract}
\maketitle





\section{Introduction}

Time evolution in standard quantum mechanics (QM) is governed by linear
equations of motion. Some outstanding issues in QM have motivated the study of
a variety of nonlinear extensions to the standard formulation of QM
\cite{Weinberg_61,Doebner_3764,Gisin_5677,Gisin_2259,Kaplan_055002,Munoz_110503,Geller_2200156,Ghirardi_470}%
. To address the problem of quantum measurement
\cite{Schrodinger_807,Penrose_4864,Oppenheim_041040}, nonlinear extensions
that give rise to spontaneous collapse of the state vector have been proposed
\cite{Bassi_471,Pearle_857,Bassi_257,Kowalski_1,Fernengel_385701}. Moreover, multistability in finite systems, which is excluded by standard QM
\cite{Chomaz_68,mainwood2005phase,Callender_539,Liu_S92,Ardourel_99,Shech_1170}%
, can be theoretically accounted for, provided that nonlinearity is permitted.

The current study explores some possible nonlinear extensions that give rise
to disentanglement \cite{Buks_2400036}. The hypothesis that disentanglement
spontaneously occurs in quantum systems is motivated by both the problem of
quantum measurement and by the difficultly to account for multi--stabilities
in standard QM \cite{Buks_012439}. Several different possibilities to
implement the hypothesis are introduced and compared below.

Consider a modified Schr\"{o}dinger equation having a form given by
\cite{Grimaudo_033835,Kowalski_167955}%
\begin{equation}
\frac{\mathrm{d}}{\mathrm{d}t}\left\vert \psi\right\rangle =\left[
-i\hbar^{-1}\mathcal{H}-\left(  \Theta-\frac{\left\langle \psi\right\vert
\Theta\left\vert \psi\right\rangle }{\left\langle \psi\right.  \left\vert
\psi\right\rangle }\right)  \right]  \left\vert \psi\right\rangle \;,
\label{MSE Theta}%
\end{equation}
where $\hbar$ is the Planck's constant, and $\mathcal{H}$ is the Hamiltonian.
The nonlinear extension in Eq. (\ref{MSE Theta}) depends on the Hermitian
operator $\Theta$, which is allowed to be dependent on the state vector
$\left\vert \psi\right\rangle $. The corresponding master equation for the
pure state density operator $\rho=\left\vert \psi\right\rangle \left\langle
\psi\right\vert $ is given by
\cite{Elben_200501,Sergi_1350163,Brody_230405,Kaplan_055002,Geller_2200156}%
\begin{equation}
\frac{\mathrm{d}\rho}{\mathrm{d}t}=i\hbar^{-1}\left[  \rho,\mathcal{H}\right]
-\Theta\rho-\rho\Theta+2\left\langle \Theta\right\rangle \frac{\rho
}{\operatorname{Tr}\rho}\;, \label{MME Theta}%
\end{equation}
where $\left\langle \Theta\right\rangle =\operatorname{Tr}\left(  \Theta
\rho\right)  $. Norm is conserved by both the modified Schr\"{o}dinger
equation (\ref{MSE Theta}) and the modified master equation (\ref{MME Theta})
[note that Eq. (\ref{MSE Theta}) yields $\left(  \mathrm{d}/\mathrm{d}%
t\right)  \left\langle \psi\right.  \left\vert \psi\right\rangle =0$, and Eq.
(\ref{MME Theta}) yields $\left(  \mathrm{d}/\mathrm{d}t\right)
\operatorname{Tr}\rho=0$]. Moreover, positivity of the density matrix $\rho$
is conserved by the modified master equation (\ref{MME Theta}) [see Eq.
(2.201) of Ref. \cite{Buks_QMLN}]. In the current study, both modified
Schr\"{o}dinger equation (\ref{MSE Theta}) and modified master equation
(\ref{MME Theta}) are employed to explore the process of disentanglement
\cite{Buks_2400587}.

The time evolution generated by the nonlinear master equation
(\ref{MME Theta}) can be expressed as (it is assumed that $\operatorname{Tr}%
\rho=1$)%
\begin{equation}
\rho\left(  t+\tau\right)  =\sum_{k\in\left\{  0,1\right\}  }K_{k}^{{}}%
\rho\left(  t\right)  K_{k}^{\dag}+O\left(  \tau^{2}\right)  \;,
\end{equation}
where the Kraus operators $K_{0}^{{}}$ and $K_{1}^{{}}$, which are given by
$K_{0}^{{}%
}=1-\left(  i\hbar^{-1}\mathcal{H}+\Theta\right)  \tau$ and $K_{1}^{{}}=\sqrt{2\left\langle \Theta\right\rangle \tau}$, satisfy the norm
conservation condition $\left\langle K_{0}^{\dag}K_{0}^{{}}+K_{1}^{\dag}%
K_{1}^{{}}\right\rangle =1+O\left(  \tau^{2}\right)  $. Note that, similarly
to the process of dephasing of standard QM [e.g. see Eq. (17) of Ref.
\cite{Daraban_048}], for the case $\mathcal{H}=0$, the Kraus operators
$K_{0}^{{}}$ and $K_{1}^{{}}$ are both Hermitian. However, while standard
dephasing occurs in a fixed basis, which is determined by the coupling
between a quantum system and its environment, the basis associated
with the nonlinear master equation (\ref{MSE Theta}), which is made of eigenvectors
of the state--dependent operator $\Theta$, is not fixed.

The nonlinear extension [in both the modified Schr\"{o}dinger equation
(\ref{MSE Theta}) and the modified master equation (\ref{MME Theta})] can be
employed to suppress any given physical property, provided that $\left\langle
\Theta\right\rangle $ quantifies that property. For example, thermalization
can be introduced by taking $\Theta$ to be given by $\Theta=\gamma
_{\mathrm{H}}\beta\mathcal{U}_{\mathrm{H}}$, where $\mathcal{U}_{\mathrm{H}%
}=\mathcal{H}+\beta^{-1}\log\rho$ is the Helmholtz free energy operator, the
real parameter $\gamma_{\mathrm{H}}$\ represents the rate of thermalization,
$\beta=1/\left(  k_{\mathrm{B}}T\right)  $ is the thermal energy inverse,
$k_{\mathrm{B}}$ is the Boltzmann's constant, and $T$ is the temperature
\cite{Grabert_161,Ottinger_052119}. The thermalization process can be
described in terms of the normalized rank of the density matrix $\rho$. For a
general $D\times D$ positive semi--definite (PSD) matrix $A$, the normalized
rank $\mathcal{R}\left(  A\right)  $ of $A$ is defined by \cite{Roy_606}%
\begin{equation}
\mathcal{R}\left(  A\right)  =\frac{1}{\log D}\operatorname{Tr}\left(
-\frac{A}{\operatorname{Tr}A}\log\frac{A}{\operatorname{Tr}A}\right)  \;.
\label{normalized rank}%
\end{equation}
Note that the normalized rank $\mathcal{R}\left(  A\right)  $ is generally
bounded by $\mathcal{R}\left(  A\right)  \in\left[  0,1\right]  $. The entropy
expectation value $\left\langle -\log\rho\right\rangle =\operatorname{Tr}%
\left(  -\rho\log\rho\right)  $ is related to the normalized rank
$\mathcal{R}\left(  \rho\right)  $ of the density
matrix $\rho$ by $\left\langle -\log\rho\right\rangle =\log\left(  D\right)
\mathcal{R}\left(  \rho\right)  $, where $D$ is the Hilbert space
dimensionality (which is assumed to be finite). As is discussed below, in a
similar way, disentanglement can be described as a matrix deranking process.

Disentanglement can be introduced provided that $\left\langle \Theta
\right\rangle $ quantifies the level of entanglement associated with the state
vector $\left\vert \psi\right\rangle $ and/or the density operator $\rho$
\cite{Schlienz_4396,Peres_1413,Hill_5022,Wootters_1717,Coffman_052306,Vedral_2275,Eltschka_424005,Dur_062314,Coiteux_200401,Takou_011004,Elben_200501,Johnston_2507_10327,Ramachandran_246,Santra_022434,Ma_1}%
. Two different types of nonlinear disentanglement operators $\Theta$ are
explored in the current study. Operators $\Theta$ belonging to the first type
give rise to matrix deranking, whereas correlation between subsystems is
suppressed by operators $\Theta$ belonging to the second type
\cite{Buks_2400587} (see appendix \ref{AppDO}). Two alternative ways, which
are based on deranking (i.e. suppression of matrix normalized rank) are
proposed and explored below. Deranking of the state matrix is discussed in
appendix \ref{AppSMEE}, whereas appendix \ref{AppBM} is devoted to deranking
of the Bloch matrix \cite{Gamel_062320}. For a pure state, both deranking
methods are applicable, whereas only Bloch matrix deranking is applicable for
general (i.e. mixed) states. Damping is accounted for using both the
Gorini-Kossakowski-Sudarshan-Lindblad (GKSL) master equation
\cite{Lindblad_119} (see appendix \ref{AppDamping}), and the
Schr\"{o}dinger--Langevin equation \cite{Jacobs_279,Zhou_272,Liu_024303} (see
appendix \ref{AppSLE}). All proposed methods to introduce disentanglement are
explored using a two spin system, which is driven close to the Hartmann--Hahn
double resonance \cite{Hartmann1962}, and which is described in the next section.

\section{Two-spin system}

\begin{figure*}[ptb]
\begin{center}
\includegraphics[width=6.4in,keepaspectratio]{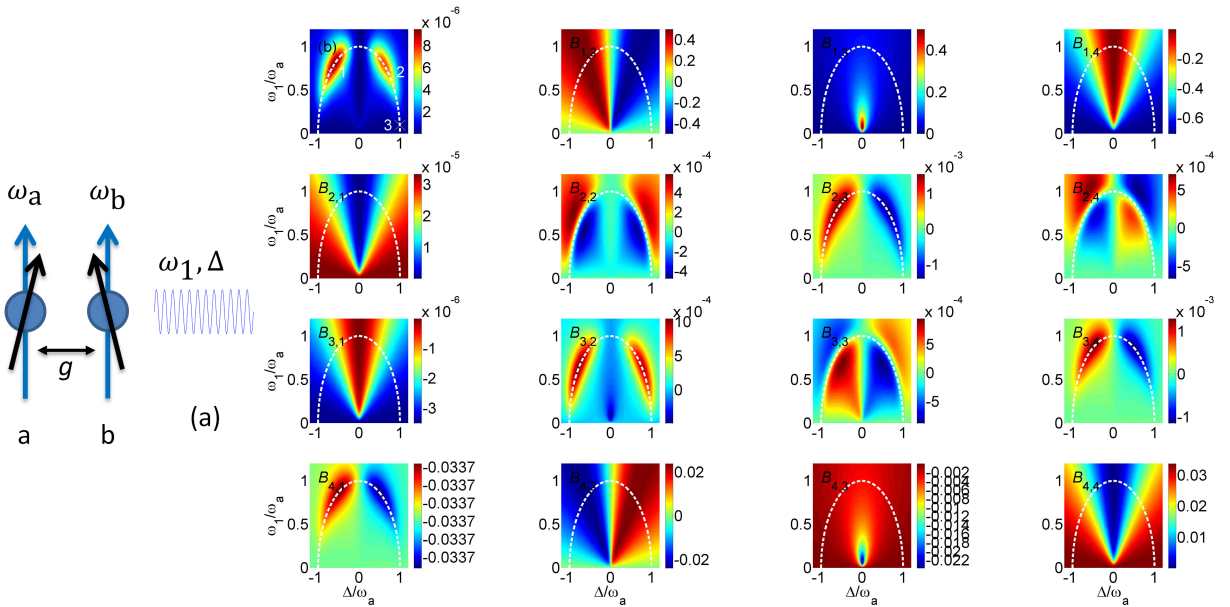}
\end{center}
\caption{{}Driving parameters. (a) A sketch of the two spin system. Steady
state value of $\tau_{\mathrm{ab}}$ is shown in (b), whereas the other color
coded plots display the steady state value of 15 Bloch matrix $B$ elements
($B_{1,1}$, which is a constant by definition, is not shown). Assumed parameters' values are $\gamma_{\mathrm{D}}=0$ (i.e. no
disentanglement), $g/\omega_{\mathrm{a}}=10^{-3}$, $\Gamma_{1}^{\left(
\mathrm{a}\right)  }/g=10$, $\Gamma_{\varphi}^{\left(  \mathrm{a}\right)
}/\Gamma_{1}^{\left(  \mathrm{a}\right)  }=10^{-4}$, $\Gamma_{1}^{\left(
\mathrm{b}\right)  }/\Gamma_{1}^{\left(  \mathrm{a}\right)  }=10$,
$\Gamma_{\varphi}^{\left(  \mathrm{b}\right)  }/\Gamma_{\varphi}^{\left(
\mathrm{a}\right)  }=10$, $\hat{n}_{0}^{\left(  \mathrm{a}\right)  }=10$ and
$\hat{n}_{0}^{\left(  \mathrm{b}\right)  }=10^{-4}$. The overlaid white
$\times$ symbols in (b) represent the assumed driving parameters (detuning
$\Delta$ and amplitude $\omega_{1}$) for the plots shown in Fig.
\ref{FigMME}.}%
\label{FigDP}%
\end{figure*}

Consider a system composed of two spin 1/2 particles [see Fig. \ref{FigDP}%
(a)]. The first spin, which is labelled as '$\mathrm{a}$', has a relatively
low Larmor angular frequency $\omega_{\mathrm{a}}$ in comparison with the
Larmor angular frequency $\omega_{\mathrm{b}}$ of the second spin, which is
labelled as '$\mathrm{b}$', and which is externally driven. The angular
momentum vector operator of spin $\mathrm{a}$ ($\mathrm{b}$) is denoted by
$\mathbf{S}_{\mathrm{a}}$ ($\mathbf{S}_{\mathrm{b}}$). The Hamiltonian
$\mathcal{H}$ of the closed two-spin system is given by%
\begin{equation}
\mathcal{H}=\omega_{\mathrm{a}}S_{\mathrm{az}}+\omega_{\mathrm{b}%
}S_{\mathrm{bz}}+\frac{\omega_{1}\left(  S_{\mathrm{b+}}+S_{\mathrm{b-}%
}\right)  }{2}+V\;, \label{H 2S}%
\end{equation}
where the driving amplitude and angular frequency are denoted by $\omega_{1}$
and $\omega_{\mathrm{p}}=\omega_{\mathrm{b}}+\Delta$, respectively ($\Delta$
is the driving detuning angular frequency), the operators $S_{\mathrm{a\pm}}$
are given by $S_{\mathrm{a\pm}}=S_{\mathrm{ax}}\pm iS_{\mathrm{ay}}$, and the
rotated operators $S_{\mathrm{b\pm}}$ are given by $S_{\mathrm{b\pm}}=\left(
S_{\mathrm{bx}}\pm iS_{\mathrm{by}}\right)  e^{\pm i\omega_{\mathrm{p}}t}$.
The dipolar coupling term $V$ is given by%
\begin{equation}
V=g\hbar^{-1}\left(  S_{\mathrm{a+}}+S_{\mathrm{a-}}\right)  S_{\mathrm{bz}%
}\;, \label{V 2S}%
\end{equation}
where $g$ is a coupling rate.

First, the case $\Theta=0$ is considered [see Eq. (\ref{MME Theta})]. Damping
is taken into account by adding a Lindblad superoperator term $\mathcal{L}$
\cite{Lindblad_119} to the master equation [see Eq. (\ref{GKSL}) of appendix
\ref{AppDamping}]. The superoperator $\mathcal{L}$ depends on the thermal
occupation factors $\hat{n}_{0\mathrm{a}}$ and $\hat{n}_{0\mathrm{b}}$, the
longitudinal relaxation times $T_{1\mathrm{a}}$ and $T_{1\mathrm{b}}$, and the
transverse relaxation times $T_{2\mathrm{a}}$ and $T_{2\mathrm{b}}$, of spins
a and b, respectively. As can be seen from Eq. (\ref{Lindbladian}) of appendix
\ref{AppDamping}, generally the Lindblad superoperator $\mathcal{L}$\ linearly
depends on the density matrix $\rho$, and consequently the master equation
(\ref{GKSL}) yields a unique steady state solution for the $4\times4$ complex
and Hermitian matrix $\rho$ (recall that in this section it is assumed that
$\Theta=0$). The corresponding $4\times4$ real Bloch matrix is denoted by $B$
[see Eq. (\ref{Bloch matrix}) of appendix \ref{AppBM}]. The dependency of $15$
(out of $16$) matrix elements of $B$ on driving detuning $\Delta$ and driving
amplitude $\omega_{1}$ is shown in Fig. \ref{FigDP} (the matrix element
$B_{1,1}$, which is a constant by definition, is not shown). The plot in Fig.
\ref{FigDP}(b) displays the correlation parameter $\tau_{\mathrm{ab}}$ [see
Eq. (\ref{tau_ab}) of appendix \ref{AppDO}] as a function of driving
parameters. As can be seen from the plots in Fig. \ref{FigDP}, the largest
effect of dipolar coupling occurs when the Hartmann--Hahn matching condition
$\omega_{\mathrm{a}}=\omega_{\mathrm{R}}$ is satisfied, where $\omega
_{\mathrm{R}}=\sqrt{\omega_{1}^{2}+\Delta^{2}}$ is the Rabi angular frequency
\cite{Hartmann1962,Yang_1,Mollow_2217}. This matching condition is represented
by the overlaid dashed white lines in the color coded plots in Fig.
\ref{FigDP}. Assumed parameters' values are listed in the caption of Fig.
\ref{FigDP}.

The first column of the matrix $B$ yields spin a Bloch vector $\mathbf{k}%
_{\mathrm{a}}=\left(  B_{2,1},B_{3,1},B_{4,1}\right)  $, whereas spin b Bloch
vector $\mathbf{k}_{\mathrm{b}}=\left(  B_{1,2},B_{1,3},B_{1,4}\right)  $ is
extracted from the first row of $B$ (recall that $B_{1,1}$ is a constant, see
appendix \ref{AppBM}). The remaining 9 elements of the Bloch
matrix $B$ represent two--spin expectation values (e.g. $B_{4,4}$ is
proportional to the expectation value $\left\langle S_{\mathrm{az}%
}S_{\mathrm{bz}} \right\rangle $).

For the case $g=0$ (i.e. no dipolar coupling), in
steady state the undriven spin a is in thermal equilibrium, and $\mathbf{k}%
_{\mathrm{a}}=\left(  0,0,k_{\mathrm{a},z}\right)  $, where $k_{\mathrm{a}%
,z}=-1/\left(  2\hat{n}_{0\mathrm{a}}+1\right)  =-\tanh\left(  \beta
\hbar\omega_{\mathrm{a}}/2\right)  $ and $\beta=1/\left(  k_{\mathrm{B}%
}T\right)  $ [see Eqs. (\ref{<S>}) and (\ref{k_z0}) of appendix
\ref{AppDamping}]. For a finite coupling coefficient $g$, the undriven spin a
in steady state is generally not in thermal equilibrium. However, an effective
temperature $T_{\mathrm{eff}}$ can be defined based on the steady state value
of $k_{\mathrm{a},z}=B_{4,1}$%
\begin{equation}
T_{\mathrm{eff}}=\frac{\hbar\omega_{\mathrm{a}}}{2k_{\mathrm{B}}\tanh
^{-1}k_{\mathrm{a},z}}\;. \label{T_eff}%
\end{equation}
The steady state value of $k_{\mathrm{a},z}=B_{4,1}$ is shown in Fig.
\ref{FigDP} as a function of spin b driving detuning $\Delta$ and driving
amplitude $\omega_{1}$. The plot of $B_{4,1}$ (bottom left subplot
in Fig. \ref{FigDP}) reveals that $T_{\mathrm{eff}}>T$ for
$\Delta<0$, and $T_{\mathrm{eff}}<T$ for $\Delta>0$, thus heating occurs with
red--detuned driving (i.e. $\Delta<0$) and cooling with blue--detuned driving
(i.e. $\Delta>0$). Note that similar driving--induced heating and cooling
effects are observed with optomechanical cavities \cite{Aspelmeyer_1391}.

As can be seen from Fig. \ref{FigDP}(b), when the Hartmann--Hahn matching
condition $\omega_{\mathrm{a}}=\omega_{\mathrm{R}}$ is satisfied (see the
overlaid white dashed line), the correlation parameter $\tau_{\mathrm{ab}}$,
which quantifies the level of entanglement between the two spins (see appendix
\ref{AppDO}), becomes relatively large. This observation suggests that the
effect of disentanglement is expected to be relatively strong for
$\omega_{\mathrm{a}} \simeq\omega_{\mathrm{R}}$.

\section{Density matrix disentanglement}

\begin{figure*}[ptb]
\begin{center}
\includegraphics[width=4in,keepaspectratio]{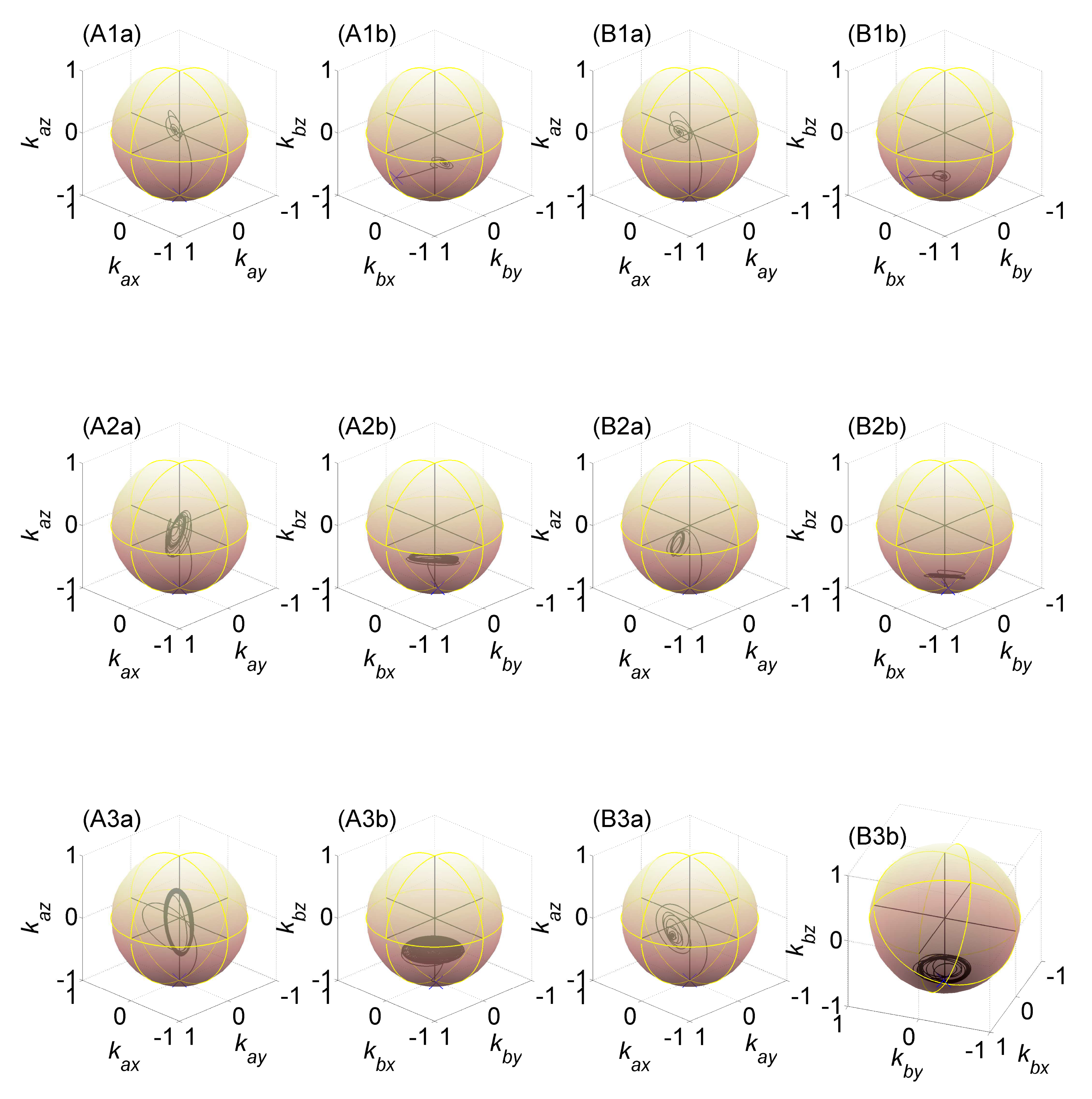}
\end{center}
\caption{{}Density matrix disentanglement. The capital letters (A
and B) in the subplots' labeling indicate the method to construct the operator
$\Theta$ (A for the case $\Theta=\gamma_{\mathrm{D}}\mathcal{Q}_{\mathrm{ab}%
}^{\left(  \mathrm{D}\right)  }$, and B for the case $\Theta=\gamma
_{\mathrm{D}}Q_{\mathrm{a}}$), the numbers (1, 2 and 3) indicate the driving
parameters [see Fig. \ref{FigDP}(b)], and the lower-case letters (a and b)
indicate the spin label. Time evolution of the single spin Bloch
vectors $\mathbf{k}_{\mathrm{a}}$ and $\mathbf{k}_{\mathrm{b}}$ is evaluated
by numerically integrating the modified master equation (\ref{MME Theta}). The
blue $\times$ symbol represents initial state, which is determined from the
steady state solution of the modified master equation (\ref{MME Theta}) for
the case $\Theta=0$. Assumed parameters' values are $g/\omega_{\mathrm{a}}=1$,
$\Gamma_{1}^{\left(  \mathrm{a}\right)  }/\omega_{\mathrm{a}}=0.1$,
$\Gamma_{\varphi}^{\left(  \mathrm{a}\right)  }/\Gamma_{1}^{\left(
\mathrm{a}\right)  }=10^{-1}$, $\Gamma_{1}^{\left(  \mathrm{b}\right)
}/\Gamma_{1}^{\left(  \mathrm{a}\right)  }=10$, $\Gamma_{\varphi}^{\left(
\mathrm{b}\right)  }/\Gamma_{\varphi}^{\left(  \mathrm{a}\right)  }=10$,
$\hat{n}_{0}^{\left(  \mathrm{a}\right)  }=5\times10^{-4}$, $\hat{n}%
_{0}^{\left(  \mathrm{b}\right)  }=1\times10^{-5}$ and $\gamma_{\mathrm{D}%
}/\omega_{\mathrm{a}}=0.5$.}%
\label{FigMME}%
\end{figure*}

The nonlinear term in the modified master equation (\ref{MME Theta}) gives
rise to disentanglement, provided that the expectation value $\left\langle
\Theta\right\rangle $ of the $\rho$--dependent operator $\Theta$ quantifies
the level of entanglement. The plots in Fig. \ref{FigMME} display time
evolution of the single spin Bloch vectors $\mathbf{k}_{\mathrm{a}}=\left(
B_{2,1},B_{3,1},B_{4,1}\right)  $ and $\mathbf{k}_{\mathrm{b}}=\left(
B_{1,2},B_{1,3},B_{1,4}\right)  $, which are calculated by numerically
integrating the modified master equation (\ref{MME Theta}).

Two different methods to construct the operator $\Theta$ are employed for
producing the plots in Fig. \ref{FigMME}. For the plots labeled by the
upper-case letter A, the operator $\Theta$ is given by $\Theta=\gamma
_{\mathrm{D}}\mathcal{Q}_{\mathrm{ab}}^{\left(  \mathrm{D}\right)  }$, whereas
$\Theta=\gamma_{\mathrm{D}}Q_{\mathrm{a}}$ for plots labeled by the upper-case
letter B. For both cases, the rate of disentanglement is denoted by
$\gamma_{\mathrm{D}}$. The operator $\mathcal{Q}_{\mathrm{ab}}^{\left(
\mathrm{D}\right)  }$ [see Eq. (\ref{Q_12 Tr}) of appendix \ref{AppDO}] gives
rise to suppression of correlation between the two spins, whereas the operator
$Q_{\mathrm{a}}$ [see Eq. (\ref{Q_a}) of appendix \ref{AppBM}] generates Bloch
matrix deranking. The labeling numbers 1, 2 and 3, which indicate the driving
parameters $\Delta/\omega_{\mathrm{a}}$ and $\omega_{1}/\omega_{\mathrm{a}}$,
refer to the overlaid white $\times$ symbols in Fig.
\ref{FigDP}(b). The lower-case letters a and b indicate the spin label.
Assumed parameters' values are listed in the caption of Fig. \ref{FigMME}.

As can be seen from Fig. \ref{FigMME}, for both cases $\Theta=\gamma
_{\mathrm{D}}\mathcal{Q}_{\mathrm{ab}}^{\left(  \mathrm{D}\right)  }$ and
$\Theta=\gamma_{\mathrm{D}}Q_{\mathrm{a}}$, with red--detuned driving the
steady state is a fixed point [see the plots of Fig. \ref{FigMME} labeled by
the number 1, and see Fig. \ref{FigDP}(b)], whereas a limit cycle steady state
can occur with blue--detuned driving [see the plots of Fig. \ref{FigMME}
labeled by the numbers 2 and 3]. The plots in Fig. \ref{FigMME} also
demonstrate that, even with the same value of the rate $\gamma_{\mathrm{D}}$,
the time evolutions generated by the operators $\mathcal{Q}_{\mathrm{ab}%
}^{\left(  \mathrm{D}\right)  }$ and $Q_{\mathrm{a}}$ are clearly distinguishable.

\section{State vector disentanglement}

In the previous section, damping was taken into account using a deterministic
master equation for the density operator $\rho$. Alternatively, damping can be
accounted for using the stochastic Schr\"{o}dinger--Langevin equation
\cite{Jacobs_279,Zhou_272,Liu_024303} for the state vector $\left\vert
\psi\right\rangle $ [see Eq. (\ref{SLE}) of appendix \ref{AppSLE}].
Disentanglement can be implemented by adding a $\Theta$--dependent nonlinear
term [see Eq. (\ref{MSE Theta})].

\begin{figure}[ptb]
\begin{center}
\includegraphics[width=3.2in,keepaspectratio]{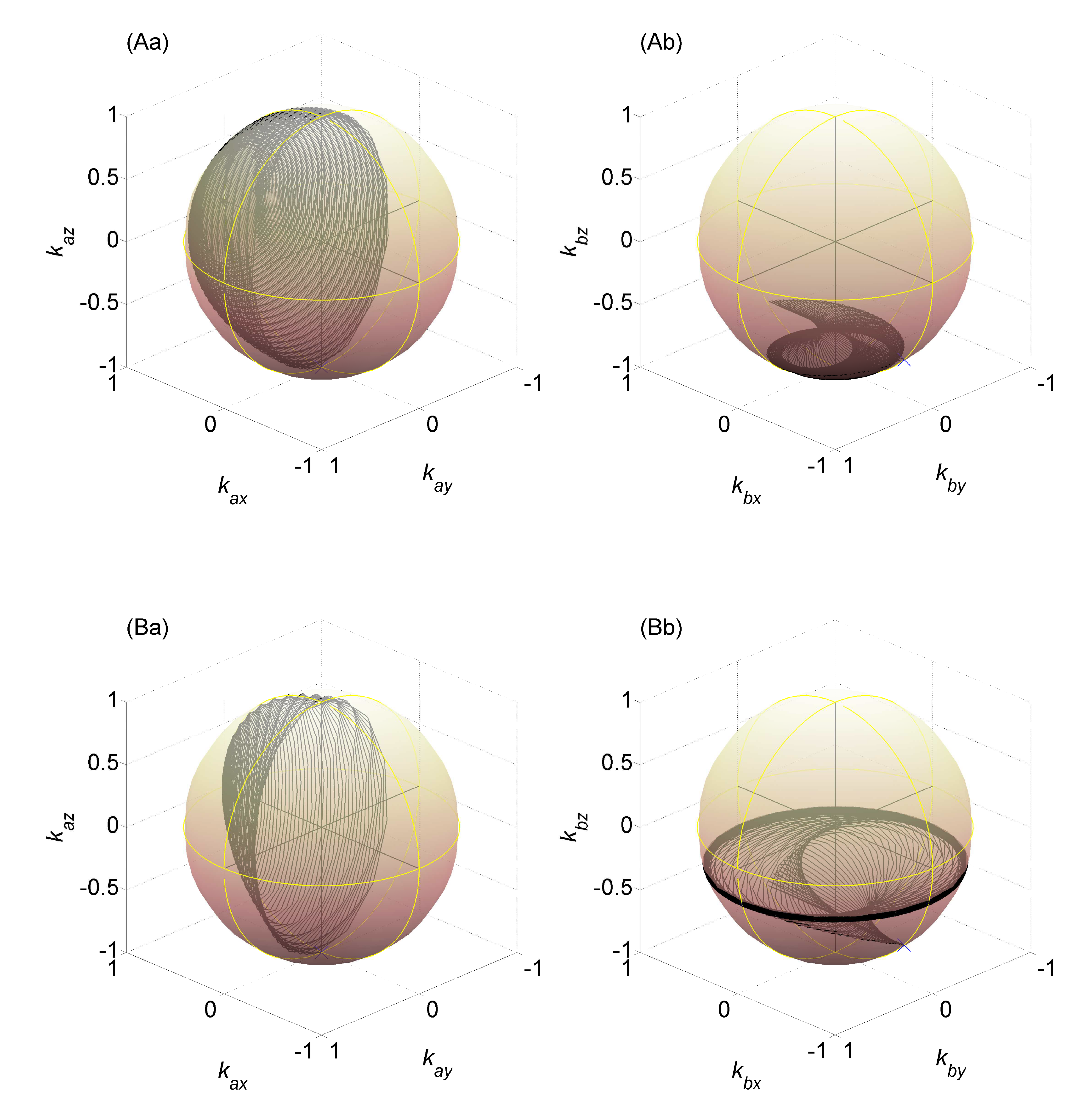}
\end{center}
\caption{{}State vector disentanglement. Time evolution of the single spin
Bloch vectors $\mathbf{k}_{\mathrm{a}}$ and $\mathbf{k}_{\mathrm{b}}$ is
evaluated by numerically integrating the modified Schr\"{o}dinger--Langevin
equation [see Eqs. (\ref{MSE Theta}) and (\ref{SLE})]. Assumed parameters'
values are $\Delta/\omega_{\mathrm{a}}=\omega_{1}/\omega_{\mathrm{a}}%
=1/\sqrt{2}$ [these driving parameters correspond to the point labeled by the
number 2 in Fig. \ref{FigDP}(b)], $g/\omega_{\mathrm{a}}=100$, $\Gamma
_{1}^{\left(  \mathrm{a}\right)  }/\omega_{\mathrm{a}}=10^{-3}$,
$\Gamma_{\varphi}^{\left(  \mathrm{a}\right)  }/\Gamma_{1}^{\left(
\mathrm{a}\right)  }=0.1$, $\Gamma_{1}^{\left(  \mathrm{b}\right)  }%
/\Gamma_{1}^{\left(  \mathrm{a}\right)  }=10$, $\Gamma_{\varphi}^{\left(
\mathrm{b}\right)  }/\Gamma_{\varphi}^{\left(  \mathrm{a}\right)  }=10$,
$\hat{n}_{0}^{\left(  \mathrm{a}\right)  }=5\times10^{-4}$ and $\hat{n}%
_{0}^{\left(  \mathrm{b}\right)  }=1\times10^{-5}$. For plots labeled by
upper-case letters A and B, $\gamma_{\mathrm{D}}/\omega_{\mathrm{a}}=0.1$ and
$\gamma_{\mathrm{D}}/\omega_{\mathrm{a}}=0.5$ , respectively.}%
\label{FigSL}%
\end{figure}

The plots in Fig. \ref{FigSL} display time evolution of the single spin Bloch
vectors $\mathbf{k}_{\mathrm{a}}$ and $\mathbf{k}_{\mathrm{b}}$, which are
calculated by numerically integrating the modified Schr\"{o}dinger--Langevin
stochastic equation [which is constructed using Eqs. (\ref{MSE Theta}) and
(\ref{SLE})]. The rate of disentanglement $\gamma_{\mathrm{D}}$ is
given by $\gamma_{\mathrm{D}}/\omega_{\mathrm{a}}=0.1$ ($\gamma_{\mathrm{D}%
}/\omega_{\mathrm{a}}=0.5$) for the plots labeled by upper-case letter A (B). The lower-case letters a and b indicate the spin label. Assumed
parameters' values are listed in the caption of Fig. \ref{FigSL}. The plots in
both Figs. \ref{FigMME} and \ref{FigSL} demonstrate the richness of dynamical
effects that can be generated by models based on the spontaneous
disentanglement hypothesis.

\section{Discussion}

The results that are presented in Figs. \ref{FigDP}, \ref{FigMME} and
\ref{FigSL} are all based on numerical calculations. For sufficiently simple
cases, analytical results, which can provide further insight, can be derived.
In particular, for the two spin system under study in a pure state, the
deranking operator $Q_{\mathrm{S}}$ [see Eq. (\ref{SM Q_S}) of appendix
\ref{AppSMEE}] is compared below to the correlation suppression operator
$\mathcal{Q}_{\mathrm{ab}}^{\left(  \mathrm{D}\right)  }$ [see Eq.
(\ref{Q_12 Tr}) of appendix \ref{AppDO}]. For a $2\times2$ state matrix $M$
given by [see Eq. (\ref{state matrix}) of appendix \ref{AppSMEE}]%
\begin{equation}
M=\left(
\begin{array}
[c]{cc}%
\psi_{1} & \psi_{2}\\
\psi_{3} & \psi_{4}%
\end{array}
\right)  \;,
\end{equation}
the PSD matrix $\mathcal{G}=M^{{}}M^{\dag}$ is [see Eq. (\ref{g matrix}) of
appendix \ref{AppSMEE}]%
\begin{equation}
\mathcal{G}=\left(
\begin{array}
[c]{cc}%
\psi_{1}\psi_{1}^{\ast}+\psi_{2}\psi_{2}^{\ast} & \psi_{1}\psi_{3}^{\ast}%
+\psi_{2}\psi_{4}^{\ast}\\
\psi_{3}\psi_{1}^{\ast}+\psi_{4}\psi_{2}^{\ast} & \psi_{3}\psi_{3}^{\ast}%
+\psi_{4}\psi_{4}^{\ast}%
\end{array}
\right)  \;.
\end{equation}
For this example $\det\mathcal{G}=\left\vert \det M\right\vert ^{2}=\left\vert
\psi_{1}\psi_{4}-\psi_{2}\psi_{3}\right\vert ^{2}=\delta/4$, where $\delta$,
which is defined by%
\begin{equation}
\delta=4\left\vert \psi_{1}\psi_{4}-\psi_{2}\psi_{3}\right\vert ^{2}\;,
\label{TS delta}%
\end{equation}
is generally bounded by $\delta\in\left[  0,1\right]  $. The eigenvalues of
$\mathcal{G}$ are $\left(  1/2\right)  \left(  1\pm\sqrt{1-\delta}\right)  $.
Note that $\delta=0$ ($\delta=1$) for a fully disentangled (fully entangled)
state. For the same pure state, the expectation value $\tau_{\mathrm{ab}}$ of
the correlation suppression operator $\mathcal{Q}_{\mathrm{ab}}^{\left(
\mathrm{D}\right)  }$ is given by [see Eq. (\ref{tau_ab}) of appendix
\ref{AppDO}]%
\begin{equation}
\tau_{\mathrm{ab}}=\left\langle \mathcal{Q}_{\mathrm{ab}}^{\left(
\mathrm{D}\right)  }\right\rangle =\frac{2\delta\left(  1+\frac{\delta}%
{2}\right)  }{3}\;. \label{tau_ab delta}%
\end{equation}
Thus, both the eigenvalues of $\mathcal{G}$ [which determine the deranking
operator $Q_{\mathrm{S}}$, see Eq. (\ref{SM Q_S}) of appendix \ref{AppSMEE}]
and $\tau_{\mathrm{ab}}$ [see Eq. (\ref{tau_ab delta})] depend on $\delta$
[which is given by Eq. (\ref{TS delta})]. On the other hand, the operator
$Q_{\mathrm{S}}$ and $\mathcal{Q}_{\mathrm{ab}}^{\left(  \mathrm{D}\right)  }$
are generally not identical, and their impacts on dynamics are distinguishable.

The method of matrix deranking is being implemented in the current study to
generate disentanglement in two different ways. In the first way, deranking is
applied to the state matrix (see appendix \ref{AppSMEE}), whereas the Bloch
matrix is being deranked in the second way (see appendix \ref{AppBM}). For the
case of a pure state, Eq. (\ref{2K=L}) of appendix \ref{AppBM} reveals that
these two ways become effectively equivalent, provided that the two subsystems
that are being disentangled share the same dimensionality.

The relation $\rho=\sum_{i}p_{i}\left\vert \alpha_{i}\right\rangle
\left\langle \alpha_{i}\right\vert $ uniquely maps any given mixed state,
which is characterized by probabilities $\left\{  p_{i}\right\}  $ (where
$0\leq p_{i}\leq1$ and $\sum_{i}p_{i}=1$), and corresponding normalized state
vectors $\left\{  \left\vert \alpha_{i}\right\rangle \right\}  $, to a density
operator $\rho$. On the other hand, a given $\rho$ generally does not uniquely
determine the mixed state (i.e. the probabilities $\left\{  p_{i}\right\}  $
and the corresponding normalized state vectors $\left\{  \left\vert \alpha
_{i}\right\rangle \right\}  $). Nevertheless, in standard QM, all ensembles
initially having the same $\rho$ share the same time evolution, which is
governed by a linear master equation for $\rho$. Moreover, in standard QM, the
deterministic master equation, which governs the time evolution of $\rho$, is
mathematically equivalent to the corresponding stochastic
Schr\"{o}dinger--Langevin equation for the state vector $\left\vert
\psi\right\rangle $ \cite{Jacobs_279}. As was discussed above, disentanglement
is taken into account by adding a nonlinear term to the master equation. For a
general mixed state, however, the impact of the nonlinear term on the time
evolution of $\rho$ is generally ensemble dependent, and it cannot be
expressed as a function of $\rho$ only. This observation suggests that, for
the studying of the impact of disentanglement in the presence of environmental
damping, it is usually advisable to implement a stochastic equation of motion
for the state vector $\left\vert \psi\right\rangle $, rather than a
deterministic master equation for the density operator $\rho$.

\section{Summary}

The current study explores some methods to implement the spontaneous
disentanglement hypothesis. All methods are applicable for any physical system
whose Hilbert space has finite dimensionality. Disentanglement has no effect
on any product (i.e. disentangled) state, thus, all predictions of standard QM
are unchanged in the absence of entanglement. The spontaneous disentanglement
hypothesis is falsifiable - its predictions are distinguishable from what is
obtained from standard QM. For a multipartite system, disentanglement between
any pair of subsystems can be introduced. Disentanglement is invariant under
any subsystem unitary transformation, and it is applicable for both
distinguishable and indistinguishable particles \cite{Buks_630}. Spontaneous
disentanglement makes the collapse postulate of QM redundant.

All under--study methods to implement disentanglement require nonlinearity,
because the subset of disentangled states in a Hilbert space of a given
quantum composite system is generally not a subspace. Nonlinear effects
\cite{Dutta_050407,Chan_051601,Landa_064301}, which are arguably inconsistent
with standard QM \cite{Buks_012439}, have been experimentally observed in a
variety of small quantum systems
\cite{Lvov_233,Roch_633,Thomas_145,Trishin_236801,Blesio_045113,Yamasaki_1187,Venkataramani_445,Koppenhofer_023026}%
. Further study is needed to explore the possibility that the underlying
mechanism responsible for the observed nonlinear effects is spontaneous disentanglement.

The process of disentanglement can give rise to limit cycle steady state
solutions (see Figs. \ref{FigMME} and \ref{FigSL}), which are otherwise
theoretically excluded. Such limit cycle solutions occur above some threshold,
which depends on the rate of disentanglement $\gamma_{\mathrm{D}}$. Upper
bounds upon $\gamma_{\mathrm{D}}$ can thus be derived from experiments
studying driven spins. The results presented here provide some guidelines for
experimentally testing the spontaneous disentanglement hypothesis. Even below
the threshold, disentanglement has an impact. In particular, it affects the
asymmetry in the response between red--detuned (i.e. $\Delta<0$) and
blue--detuned (i.e. $\Delta>0$) driving. Moreover, disentanglement can give
rise to multistability, which is otherwise excluded (for sufficiently small
systems) \cite{Buks_2400587,Yin_011064}.

\appendix

\section{Damping}

\label{AppDamping}

The GKSL master equation for the reduced density operator $\rho$ is given by
\cite{Fernengel_385701,Lindblad_119,Manzano_025106}%
\begin{equation}
\frac{\mathrm{d}\rho}{\mathrm{d}t}=i\hbar^{-1}\left[  \rho,\mathcal{H}\right]
+\mathcal{L}\;, \label{GKSL}%
\end{equation}
where $\mathcal{H}^{{}}=\mathcal{H}^{\dag}$ is the Hamiltonian, and
$\mathcal{L}$ is a Lindblad superoperator \cite{Lindblad_119}. For the two
spin system under study, the coupling between spin $\mathrm{L}$ and its
environment, where $\mathrm{L}\in\left\{  \mathrm{a},\mathrm{b}\right\}  $, is
characterized by energy--relaxation $\Gamma_{1}^{\left(  \mathrm{L}\right)  }$
and dephasing $\Gamma_{\varphi}^{\left(  \mathrm{L}\right)  }$ rates, thermal
occupation factor $\hat{n}_{0}^{\left(  \mathrm{L}\right)  }$, and
longitudinal $T_{1}^{\left(  \mathrm{L}\right)  }$ and transverse
$T_{2}^{\left(  \mathrm{L}\right)  }$ relaxation times. The Lindblad
superoperator $\mathcal{L}$ is given by \cite{carmichael2009open}%
\begin{align}
\mathcal{L}  &  =\sum_{\mathrm{L}\in\left\{  \mathrm{a},\mathrm{b}\right\}
}\frac{\left(  \hat{n}_{0}^{\left(  \mathrm{L}\right)  }+1\right)  \Gamma
_{1}^{\left(  \mathrm{L}\right)  }}{4}\mathcal{D}_{\rho}\left(  \frac
{2S_{\mathrm{L},-}}{\hbar}\right) \nonumber\\
&  +\frac{\hat{n}_{0}^{\left(  \mathrm{L}\right)  }\Gamma_{1}^{\left(
\mathrm{L}\right)  }}{4}\mathcal{D}_{\rho}\left(  \frac{2S_{\mathrm{L},+}%
}{\hbar}\right) \nonumber\\
&  +\frac{\left(  2\hat{n}_{0}^{\left(  \mathrm{L}\right)  }+1\right)
\Gamma_{\varphi}^{\left(  \mathrm{L}\right)  }}{2}\mathcal{D}_{\rho}\left(
\frac{2S_{\mathrm{L},z}}{\hbar}\right)  \;,\nonumber\\
&  \label{superoperator}%
\end{align}
where the Lindbladian $\mathcal{D}_{\rho}\left(  X\right)  $ for an operator
$X$\ is given by%
\begin{equation}
\mathcal{D}_{\rho}\left(  X\right)  =X^{{}}\rho X^{\dag}-\frac{X^{\dag}X^{{}%
}\rho+\rho X^{\dag}X^{{}}}{2}\;. \label{Lindbladian}%
\end{equation}
The positive damping rates $\Gamma_{1}^{\left(  \mathrm{L}\right)  }$ and
$\Gamma_{\varphi}^{\left(  \mathrm{L}\right)  }$, and the thermal occupation
factor $\hat{n}_{0}^{\left(  \mathrm{L}\right)  }$, are related to the
longitudinal $T_{1}^{\left(  \mathrm{L}\right)  }$ and the transverse
$T_{2}^{\left(  \mathrm{L}\right)  }$ relaxation times, and to the thermal
equilibrium spin polarization $P_{z0}^{\left(  \mathrm{L}\right)  }$, by
$1/T_{1}^{\left(  \mathrm{L}\right)  }=-\Gamma_{1}^{\left(  \mathrm{L}\right)
}/P_{z0}^{\left(  \mathrm{L}\right)  }$, $1/T_{2}^{\left(  \mathrm{L}\right)
}=-\left(  \Gamma_{1}^{\left(  \mathrm{L}\right)  }/2+\Gamma_{\varphi
}^{\left(  \mathrm{L}\right)  }\right)  /P_{z0}^{\left(  \mathrm{L}\right)  }$
and $-1/P_{z0}^{\left(  \mathrm{L}\right)  }=2\hat{n}_{0}^{\left(
\mathrm{L}\right)  }+1$.

As an example, consider a single spin 1/2 under transverse driving having
amplitude $\omega_{1}$ and angular frequency $\omega_{\mathrm{p}}%
=\omega_{\mathrm{L}}+\Delta$, where $\omega_{\mathrm{L}}$ is the Larmor
angular frequency, and $\Delta$ is the driving angular frequency detuning [see
Eq. (\ref{H 2S}) for $\omega_{\mathrm{a}}=0$ and $V=0$]. For that case, the
GKSL master equation (\ref{GKSL}) yields a unique steady state solution, for
which the expectation value of the spin angular momentum vector operator
$\left\langle \mathbf{S}\right\rangle $ is given by \cite{carmichael2009open}%
\begin{equation}
\frac{2}{\hbar}\left\langle \mathbf{S}\right\rangle =\left(
\begin{array}
[c]{c}%
\frac{\Delta\omega_{1}T_{2}^{2}P_{z0}}{1+\Delta^{2}T_{2}^{2}+\omega_{1}%
^{2}T_{1}T_{2}}\\
-\frac{\omega_{1}T_{2}P_{z0}}{1+\Delta^{2}T_{2}^{2}+\omega_{1}^{2}T_{1}T_{2}%
}\\
\frac{\left(  1+\Delta^{2}T_{2}^{2}\right)  P_{z0}}{1+\Delta^{2}T_{2}%
^{2}+\omega_{1}^{2}T_{1}T_{2}}%
\end{array}
\right)  \;, \label{<S>}%
\end{equation}
where%
\begin{equation}
P_{z0}=-\frac{1}{2\hat{n}+1}=-\tanh\frac{\beta\hbar\omega_{\mathrm{L}}}{2}\;,
\label{k_z0}%
\end{equation}
and $\beta=1/\left(  k_{\mathrm{B}}T\right)  $.

\section{State matrix and entanglement entropy}

\label{AppSMEE}

Consider a $D_{\mathrm{H}}$--dimensional Hilbert space, where $D_{\mathrm{H}%
}\in\left\{  4,5,6,\cdots\right\}  $ is finite. Any state in the Hilbert space
is represented by a complex $D_{\mathrm{H}}\times1$ column vector given by%
\begin{equation}%
\genfrac{\vert}{)}{0pt}{}{\psi}{D_{\mathrm{H}}}%
=\left(
\begin{array}
[c]{c}%
\psi_{1}\\
\psi_{2}\\
\vdots\\
\psi_{D_{\mathrm{H}}}%
\end{array}
\right)  \;.
\end{equation}
The symbols $\left\vert {}\right)  $ and $\left(  {}\right\vert $ are
henceforth used to denote column (ket) and row (bra) vectors, respectively. A
unit $N\times1$ column vector, whose $m$'th entry is given by $\delta_{m,n}$,
where $n\in\left\{  1,2,\cdots,N\right\}  $, is denoted by $%
\genfrac{\vert}{)}{0pt}{}{n}{N}%
$. For the case where the variable $x$ does not represent an
integer number, the symbol $%
\genfrac{\vert}{)}{0pt}{}{x}{N}%
$ denotes a general $N\times1$ column vector, and the symbol $%
\genfrac{(}{\vert}{0pt}{}{x}{N}%
$ denotes its Hermitian conjugate $1\times N$ row vector.

Unless $D_{\mathrm{H}}$ is prime, it can be factored as $D_{\mathrm{H}%
}=D_{\mathrm{a}}D_{\mathrm{b}}$, where $D_{\mathrm{a}}>1$ and $D_{\mathrm{b}%
}>1$ are both integers. The two subsystems corresponding to the factorization
\cite{Carroll_022213} are labelled as 'a' and 'b', respectively. For any given
factorization, the state $\psi$ is represented by a $D_{\mathrm{a}}\times
D_{\mathrm{b}}$ \textit{state matrix} $M$ given by%
\begin{equation}
M=\left(
\begin{array}
[c]{cccc}%
\psi_{1} & \psi_{2} & \cdots & \psi_{D_{\mathrm{b}}}\\
\psi_{D_{\mathrm{b}}+1} & \psi_{D_{\mathrm{b}}+2} & \cdots & \psi
_{2D_{\mathrm{b}}}\\
\vdots & \vdots & \ddots & \vdots\\
\psi_{\left(  D_{\mathrm{a}}-1\right)  D_{\mathrm{b}}+1} & \psi_{\left(
D_{\mathrm{a}}-1\right)  D_{\mathrm{b}}+2} &  & \psi_{D_{\mathrm{a}%
}D_{\mathrm{b}}}%
\end{array}
\right)  \;. \label{state matrix}%
\end{equation}
The $D_{\mathrm{a}}\times D_{\mathrm{a}}$ PSD matrix $\mathcal{G}$ is defined
by%
\begin{equation}
\mathcal{G}=M^{{}}M^{\dag}\;. \label{g matrix}%
\end{equation}
For any eigenvalue $\lambda$ of $\mathcal{G}$, the following holds
$0\leq\lambda\leq1$, $\lambda$ is an eigenvalue of the $D_{\mathrm{b}}\times
D_{\mathrm{b}}$ PSD matrix $M^{\dag}M^{{}}$, and $\lambda^{1/2}$ is a singular
value of $M$ \cite{Parzygnat_1}. Note that the normalization condition $%
\genfrac{(}{.}{0pt}{}{\psi}{D_{\mathrm{H}}}%
\genfrac{\vert}{)}{0pt}{}{\psi}{D_{\mathrm{H}}}%
=1$ yields $\operatorname{Tr}\mathcal{G}=1$.

The level of bipartite entanglement (aka entanglement entropy) associated with
a given pure state $%
\genfrac{\vert}{)}{0pt}{}{\psi}{D_{\mathrm{H}}}%
$ and a given factorization $D_{\mathrm{H}}=D_{\mathrm{a}}D_{\mathrm{b}}$,
which is henceforth denoted by $\mathcal{K}$, can be characterized in term of
the normalized rank $\mathcal{R}\left(  \mathcal{G}\right)  $ of the
$D_{\mathrm{a}}\times D_{\mathrm{a}}$ PSD matrix $\mathcal{G}$ as [recall that
$\operatorname{Tr}\mathcal{G}=1$, and see Eq. (\ref{normalized rank})]%
\begin{equation}
\mathcal{K}=\log\left(  D_{\mathrm{a}}\right)  \mathcal{R}\left(
\mathcal{G}\right)  =\operatorname{Tr}\left(  -\mathcal{G}\log\mathcal{G}%
\right)  \;. \label{K LOE}%
\end{equation}
Note that $\mathcal{K}$ is bounded by $\mathcal{K}\in\left[  0,\log\left(
\min\left\{  D_{\mathrm{a}},D_{\mathrm{b}}\right\}  \right)  \right]  $. To
implement disentanglement by deranking, the level of bipartite entanglement
$\mathcal{K}$ has to be expressed as an expectation value. An Hermitian
operator $Q_{\mathrm{S}}$ satisfying the relation $\left\langle Q_{\mathrm{S}%
}\right\rangle =\mathcal{K}$ is derived below.

The state matrix $M$ can be expressed as%
\begin{equation}
M=\sum_{d_{\mathrm{a}}=1}^{D_{\mathrm{a}}}\sum_{d_{\mathrm{b}}=1}%
^{D_{\mathrm{b}}}\mu_{d_{\mathrm{a}},D_{\mathrm{a}},d_{\mathrm{b}%
},D_{\mathrm{b}}}%
\genfrac{\vert}{)}{0pt}{}{\psi}{D_{\mathrm{a}}D_{\mathrm{b}}}%
\genfrac{(}{\vert}{0pt}{}{d_{\mathrm{b}}}{D_{\mathrm{b}}}%
\;,
\end{equation}
where the $D_{\mathrm{a}}\times\left(  D_{\mathrm{a}}D_{\mathrm{b}}\right)  $
matrix $\mu_{d_{\mathrm{a}},D_{\mathrm{a}},d_{\mathrm{b}},D_{\mathrm{b}}}$ is
given by%
\begin{align}
\mu_{d_{\mathrm{a}},D_{\mathrm{a}},d_{\mathrm{b}},D_{\mathrm{b}}}  &  =%
\genfrac{\vert}{)}{0pt}{}{d_{\mathrm{a}}}{D_{\mathrm{a}}}%
\left(
\genfrac{(}{\vert}{0pt}{}{d_{\mathrm{a}}}{D_{\mathrm{a}}}%
\otimes%
\genfrac{(}{\vert}{0pt}{}{d_{\mathrm{b}}}{D_{\mathrm{b}}}%
\right) \nonumber\\
&  =%
\genfrac{\vert}{)}{0pt}{}{d_{\mathrm{a}}}{D_{\mathrm{a}}}%
\genfrac{(}{\vert}{0pt}{}{\left(  d_{\mathrm{a}}-1\right)  D_{\mathrm{b}%
}+d_{\mathrm{b}}}{D_{\mathrm{a}}D_{\mathrm{b}}}%
\;,\nonumber\\
&
\end{align}
and where the symbol $\otimes$ denotes Kronecker matrix product, and thus%
\begin{equation}
\mathcal{G}\equiv M^{{}}M^{\dag}=\sum_{d_{\mathrm{b}}=1}^{D_{\mathrm{b}}}%
\sum_{d_{\mathrm{a}}^{\prime}=1}^{D_{\mathrm{a}}}\sum_{d_{\mathrm{a}}%
^{\prime\prime}=1}^{D_{\mathrm{a}}}\mu_{d_{\mathrm{a}}^{\prime},D_{\mathrm{a}%
},d_{\mathrm{b}},D_{\mathrm{b}}}^{{}}\rho_{\psi}\mu_{d_{\mathrm{a}}%
^{\prime\prime},D_{\mathrm{a}},d_{\mathrm{b}},D_{\mathrm{b}}}^{\dag}\;,
\label{G SM}%
\end{equation}
where%
\begin{equation}
\rho_{\psi}=%
\genfrac{\vert}{)}{0pt}{}{\psi}{D_{\mathrm{a}}D_{\mathrm{b}}}%
\genfrac{(}{\vert}{0pt}{}{\psi}{D_{\mathrm{a}}D_{\mathrm{b}}}%
\;.
\end{equation}

The definition of the level of entanglement $\mathcal{K}$, which for a pure
state is given by Eq. (\ref{K LOE}), is generalized for a mixed state by [see
Eq. (\ref{G SM}), and recall the identity $\operatorname{Tr}\left(  \left\vert
u\right\rangle \left\langle v\right\vert \right)  =\left\langle v\right.
\left\vert u\right\rangle $]%
\begin{equation}
\mathcal{K}=\operatorname{Tr}\left(  \rho Q_{\mathrm{S}}\right)
\equiv\left\langle Q_{\mathrm{S}}\right\rangle \;, \label{K general state}%
\end{equation}
where the operator $Q_{\mathrm{S}}$ is given by%
\begin{equation}
Q_{\mathrm{S}}=-\sum_{d_{\mathrm{b}}=1}^{D_{\mathrm{b}}}\sum_{d_{\mathrm{a}%
}^{\prime}=1}^{D_{\mathrm{a}}}\sum_{d_{\mathrm{a}}^{\prime\prime}%
=1}^{D_{\mathrm{a}}}\mu_{d_{\mathrm{a}}^{\prime\prime},D_{\mathrm{a}%
},d_{\mathrm{b}},D_{\mathrm{b}}}^{\dag}\left(  \log\mathcal{G}\right)
\mu_{d_{\mathrm{a}}^{\prime},D_{\mathrm{a}},d_{\mathrm{b}},D_{\mathrm{b}}}%
^{{}}\;, \label{SM Q_S}%
\end{equation}
and where $\rho$ is the density matrix of the given state [the matrix
$\mathcal{G}$ is calculated using Eq. (\ref{G SM}) for $\rho_{\psi}=\rho$].

\section{Bloch matrix}

\label{AppBM}

For a given Hilbert space having dimensionality $D_{\mathrm{H}}$, and for a
given factorization $D_{\mathrm{H}}=D_{\mathrm{a}}D_{\mathrm{b}}$, where
$D_{\mathrm{a}}>1$ and $D_{\mathrm{b}}>1$ are both integers, a state matrix
$M$ [see Eq. (\ref{state matrix}) of appendix \ref{AppSMEE}] can be defined
for any \textit{pure} state, whereas a Bloch matrix can be defined for any
general (i.e. mixed) state. The generalized Gell-Mann set $\left\{
\lambda_{l}\right\}  $, which spans the SU($D_{\mathrm{H}}$) Lie algebra,
contains $D_{\mathrm{H}}^{2}-1$ square $D_{\mathrm{H}}\times D_{\mathrm{H}}$
Hermitian matrices. For the case $D_{\mathrm{H}}=2$ ($D_{\mathrm{H}}=3$), the
$D_{\mathrm{H}}^{2}-1=3$ ($D_{\mathrm{H}}^{2}-1=8$) elements are called Pauli
(Gell-Mann) matrices. The Generalized Gell-Mann matrices are traceless, i.e.
$\operatorname{Tr}\lambda_{l}=0$, and they satisfy the orthogonality relation%
\begin{equation}
\frac{\operatorname{Tr}\left(  \lambda_{l^{\prime}}\lambda_{l^{\prime\prime}%
}\right)  }{2}=\delta_{l^{\prime},l^{\prime\prime}}\;. \label{GM OR}%
\end{equation}

For a given factorization $D_{\mathrm{H}}=D_{\mathrm{a}}D_{\mathrm{b}}$, the
generalized Gell-Mann $D_{\mathrm{L}}\times D_{\mathrm{L}}$ matrices
corresponding to subsystem $\mathrm{L}$, where $\mathrm{L}\in\left\{
\mathrm{a},\mathrm{b}\right\}  $, are denoted by $\lambda_{l}^{\left(
\mathrm{L}\right)  }$, where $l\in\left\{  1,2,\cdots,D_{\mathrm{L}}%
^{2}-1\right\}  $. Consider the set of $D_{\mathrm{H}}^{2}-1$ matrices
$G^{\left(  \mathrm{ab}\right)  }=\left\{  \Gamma_{a}^{\left(  \mathrm{a}%
\right)  }\otimes\Gamma_{b}^{\left(  \mathrm{b}\right)  }\right\}  -\left\{
\Gamma_{0}^{\left(  \mathrm{a}\right)  }\otimes\Gamma_{0}^{\left(
\mathrm{b}\right)  }\right\}  $, where $a\in\left\{  0,1,2,\cdots
,D_{\mathrm{a}}^{2}-1\right\}  $ and $b\in\left\{  0,1,2,\cdots,D_{\mathrm{b}%
}^{2}-1\right\}  $. For subsystem $\mathrm{L}$, where $\mathrm{L}\in\left\{
\mathrm{a},\mathrm{b}\right\}  $, the matrix $\Gamma_{0}^{\left(
\mathrm{L}\right)  }$ is defined by $\Gamma_{0}^{\left(  \mathrm{L}\right)
}=\left(  2^{1/4}/D_{\mathrm{L}}^{1/2}\right)  I_{\mathrm{L}}$,\ where
$I_{\mathrm{L}}$ is the $D_{\mathrm{L}}\times D_{\mathrm{L}}$ identity matrix,
and for $l\in\left\{  1,2,\cdots,D_{\mathrm{L}}^{2}-1\right\}  $\ the matrix
$\Gamma_{l}^{\left(  \mathrm{L}\right)  }$ is defined by $\Gamma_{l}^{\left(
\mathrm{L}\right)  }=2^{-1/4}\lambda_{l}^{\left(  \mathrm{L}\right)  }$.

With the help of the Kronecker matrix product identities $\operatorname{Tr}%
\left(  X_{1}\otimes X_{2}\right)  =\operatorname{Tr}X_{1}\operatorname{Tr}%
X_{2}$ and $\left(  X_{1}\otimes X_{2}\right)  \left(  X_{3}\otimes
X_{4}\right)  =\left(  X_{1}X_{3}\right)  \otimes\left(  X_{2}X_{4}\right)  $,
one finds that the set $G^{\left(  \mathrm{ab}\right)  }$ shares two
properties with the Gell-Mann set $G$ of the $D_{\mathrm{H}}$-dimensional
Hilbert space. The first one is tracelessness $\operatorname{Tr}G_{a,b}=0$ for
any $G_{a,b}\equiv\Gamma_{a}^{\left(  \mathrm{a}\right)  }\otimes\Gamma
_{b}^{\left(  \mathrm{b}\right)  }\in G^{\left(  \mathrm{ab}\right)  }$
[recall that $G_{0,0}\notin G^{\left(  \mathrm{ab}\right)  }$], and the second
one is orthogonality [see Eq. (\ref{GM OR})]%
\begin{equation}
\frac{\operatorname{Tr}\left(  G_{a^{\prime},b^{\prime}}G_{a^{\prime\prime
},b^{\prime\prime}}\right)  }{2}=\delta_{a^{\prime},a^{\prime\prime}}%
\delta_{b^{\prime},b^{\prime\prime}}\;. \label{G OR}%
\end{equation}

The $D_{\mathrm{a}}^{2}\times D_{\mathrm{b}}^{2}$ matrix $B$, where%
\begin{equation}
B_{a,b}=\left\langle G_{a,b}\right\rangle ^{{}}\;, \label{Bloch matrix}%
\end{equation}
is henceforth referred to as the Bloch matrix. The following holds
$B_{0,0}=\sqrt{2/\left(  D_{\mathrm{a}}D_{\mathrm{b}}\right)  }$, and
$\operatorname{Tr}\left(  B^{{}}B^{\dag}\right)  =\operatorname{Tr}\left(
B^{\dag}B^{{}}\right)  =2\operatorname{Tr}\rho^{2}$ [see Eq. (\ref{G OR}), and
recall the identity $\operatorname{Tr}\left(  X_{1}\otimes X_{2}\right)
=\operatorname{Tr}X_{1}\operatorname{Tr}X_{2}$]. Expectation value
$\left\langle A\right\rangle =\operatorname{Tr}\left(  \rho A\right)  $ of a
given observable $A$ is given by%
\begin{equation}
\left\langle A\right\rangle =\sum_{a=0}^{D_{\mathrm{a}}^{2}-1}\sum
_{b=0}^{D_{\mathrm{b}}^{2}-1}\frac{\operatorname{Tr}\left(  \rho
G_{a,b}\right)  \operatorname{Tr}\left(  AG_{a,b}\right)  }{2}\;.
\end{equation}

The Bloch matrix $B$ can be used to define an alternative quantification for
the level of bipartite entanglement $\mathcal{L}$, which is given by [compare
with Eq. (\ref{K LOE})]%
\begin{equation}
\mathcal{L}=-\operatorname{Tr}\left(  \alpha\log\alpha\right)
=-\operatorname{Tr}\left(  \beta\log\beta\right)  \;, \label{L LOE}%
\end{equation}
where the $D_{\mathrm{a}}^{2}\times D_{\mathrm{a}}^{2}$ matrix $\alpha$ is
given by $\alpha=\left(  1/2\right)  B^{{}}B^{\dag}$, and the $D_{\mathrm{b}%
}^{2}\times D_{\mathrm{b}}^{2}$ matrix $\beta$ is given by $\beta=\left(
1/2\right)  B^{\dag}B^{{}}$ (recall that the PSD matrices $\alpha$ and $\beta$
share the same set of eigenvalues, and the same trace $\operatorname{Tr}%
\alpha=\operatorname{Tr}\beta=\operatorname{Tr}\rho^{2}$). The following holds
[see Eq. (\ref{L LOE})]%
\begin{equation}
\left\langle Q_{\mathrm{a}}\right\rangle =\left\langle Q_{\mathrm{b}%
}\right\rangle =\mathcal{L}\;, \label{<Q_a>=<Q_b>=L}%
\end{equation}
where the operators $Q_{\mathrm{a}}$ and $Q_{\mathrm{b}}$\ are defined by%
\begin{align}
Q_{\mathrm{a}}  &  =\operatorname{Tr}\left(  -\frac{G^{{}}B^{\dag}}{2}%
\log\left(  \frac{B^{{}}B^{\dag}}{2}\right)  \right)  \;,\label{Q_a}\\
Q_{\mathrm{b}}  &  =\operatorname{Tr}\left(  -\log\left(  \frac{B^{\dag}B^{{}%
}}{2}\right)  \frac{B^{\dag}G^{{}}}{2}\right)  \;. \label{Q_b}%
\end{align}
The $\left(  a,b\right)  $ entry of the $D_{\mathrm{a}}^{2}\times
D_{\mathrm{b}}^{2}$ matrix $G$ is the $D_{\mathrm{H}}\times D_{\mathrm{H}}$
observable $G_{a,b}=\Gamma_{a}^{\left(  \mathrm{a}\right)  }\otimes\Gamma
_{b}^{\left(  \mathrm{b}\right)  }$, and the $\left(  a,b\right)  $ entry of
the $D_{\mathrm{a}}^{2}\times D_{\mathrm{b}}^{2}$ matrix $B$ is the
expectation value $B_{a,b}=\left\langle G_{a,b}\right\rangle $ [note that
$a\in\left\{  0,1,\cdots,D_{\mathrm{a}}^{2}-1\right\}  $ and $b\in\left\{
0,1,\cdots,D_{\mathrm{b}}^{2}-1\right\}  $]. As can be see from Eq.
(\ref{<Q_a>=<Q_b>=L}), the operators $Q_{\mathrm{a}}$ and $Q_{\mathrm{b}}$ can
be used for the implementation of disentanglement that is based on the
deranking of the matrices $\alpha$ and $\beta$.

Consider a pure state that is characterized by a state matrix $M$ [see Eq.
(\ref{state matrix}) of appendix \ref{AppSMEE}] and a Bloch matrix $B$ [see
Eq. (\ref{Bloch matrix})]. Recall that for a pure state $\operatorname{Tr}%
\left(  \left(  1/2\right)  B^{{}}B^{\dag}\right)  =\operatorname{Tr}\left(
\left(  1/2\right)  B^{\dag}B^{{}}\right)  =\operatorname{Tr}\rho^{2}=1$. For
the case $D_{\mathrm{a}}=D_{\mathrm{b}}$, the Hermitian matrices $\left(
M^{\dag}M^{{}}\right)  \otimes\left(  M^{\dag}M^{{}}\right)  $ and $\left(
1/2\right)  B^{{}}B^{\dag}$ are unitarily equivalent [proof is based on
Specht's theorem, see Eq. (8.1046) of Ref. \cite{Buks_QMLN}]. This implies
that for this case the entanglement parameter $\mathcal{K}$ [see Eq.
(\ref{K LOE}) of appendix \ref{AppSMEE}], which is based on the state matrix
$M$ [see Eq. (\ref{state matrix}) of appendix \ref{AppSMEE}], and the
entanglement parameter $\mathcal{L}$ [see Eq. (\ref{L LOE})], which is based
on the Bloch matrix $B$ (\ref{Bloch matrix}), are related by [recall the
tensor product identity $\operatorname{Tr}\left(  X\otimes Y\right)
=\operatorname{Tr}X\operatorname{Tr}Y$]%
\begin{equation}
2\mathcal{K}=\mathcal{L}\;. \label{2K=L}%
\end{equation}

\section{Weyl basis}

\label{AppWB}

The generalized Gell-Mann matrices have been employed in appendix \ref{AppBM}
to span the SU($D_{\mathrm{H}}$) Lie algebra. Alternatively, the Weyl
operators can be used for the same purpose
\cite{Siewert_055014,Lai_115305,Huang_1064}. Let $\left\{  \left\vert
n\right\rangle \right\}  $, where $n\in\left\{  0,1,2,\cdots,D-1\right\}  $,
be an orthonormal basis for a Hilbert space of a $D$ level system. The Weyl
operator $W_{n^{\prime}n^{\prime\prime}}$ is defined by%
\begin{equation}
W_{n^{\prime}n^{\prime\prime}}=\sum_{n=0}^{D-1}e^{\frac{2\pi i}{D}n^{{}%
}n^{\prime}}\left\vert n^{{}}\right\rangle \left\langle n^{{}}+n^{\prime
\prime}\right\vert \;, \label{NLS Weyl operators}%
\end{equation}
where $n^{\prime},n^{\prime\prime}\in\left\{  0,1,2,\cdots,D-1\right\}  $. For
any integer $n$, the abbreviated notation $\left\vert n\right\rangle $ denotes
the state $\left\vert \operatorname{mod}\left(  n,D\right)  \right\rangle $.
For a mixed state represented by a $D\times D$ density matrix $\rho$, the
elements of the Weyl matrix $\mathcal{W}$ are given by (note that elements'
numbering starts from zero)%
\begin{equation}
\mathcal{W}_{n^{\prime}n^{\prime\prime}}=\frac{\operatorname{Tr}\left(
W_{n^{\prime}n^{\prime\prime}}\rho\right)  }{\sqrt{D}}\;,
\label{SNL Weyl matrix rho}%
\end{equation}
where $n^{\prime},n^{\prime\prime}\in\left\{  0,1,2,\cdots,D-1\right\}  $. For
a pure state $\operatorname{Tr}\left(  \mathcal{W}^{\dag}\mathcal{W}^{{}%
}\right)  =1$ and $1/D\leq\operatorname{Tr}\left(  \left(  \mathcal{W}^{\dag
}\mathcal{W}^{{}}\right)  ^{2}\right)  \leq1$, whereas for a mixed state
$1/D^{2}\leq\operatorname{Tr}\left(  \mathcal{W}^{\dag}\mathcal{W}^{{}%
}\right)  =\operatorname{Tr}\rho^{2}\leq1$.

Consider a system composed of two subsystems labelled by the letters a and b,
respectively. The Hilbert space dimensionality of subsystem a (b) is denoted
by $D_{\mathrm{a}}$ ($D_{\mathrm{b}}$). The Weyl operators
(\ref{NLS Weyl operators}) corresponding to subsystem $\mathrm{L}$, where
$\mathrm{L}\in\left\{  \mathrm{a},\mathrm{b}\right\}  $, are denoted by
$W_{n^{\prime}n^{\prime\prime}}^{\left(  \mathrm{L}\right)  }$, where
$n^{\prime},n^{\prime\prime}\in\left\{  0,1,2,\cdots,D_{\mathrm{L}}-1\right\}
$. For a given $D_{\mathrm{a}}D_{\mathrm{b}}\times D_{\mathrm{a}}%
D_{\mathrm{b}}$ density matrix $\rho$, the elements of the $D_{\mathrm{a}%
}D_{\mathrm{b}}\times D_{\mathrm{a}}D_{\mathrm{b}}$ Weyl matrix $\mathcal{W}$
are given by [see Eq. (\ref{SNL Weyl matrix rho})]%
\begin{equation}
\mathcal{W}_{\left(  n^{\prime},n^{\prime\prime\prime}\right)  ,\left(
n^{\prime\prime},n^{\prime\prime\prime\prime}\right)  }=\frac
{\operatorname{Tr}\left(  \left(  W_{n^{\prime}n^{\prime\prime}}^{\left(
\mathrm{a}\right)  }\otimes W_{n^{\prime\prime\prime}n^{\prime\prime
\prime\prime}}^{\left(  \mathrm{b}\right)  }\right)  \rho\right)  }%
{\sqrt{D_{\mathrm{a}}D_{\mathrm{b}}}}\;, \label{NLS TS Weyl matrix}%
\end{equation}
where $n^{\prime},n^{\prime\prime}\in\left\{  0,1,2,\cdots,D_{\mathrm{a}%
}-1\right\}  $ and $n^{\prime\prime\prime},n^{\prime\prime\prime\prime}%
\in\left\{  0,1,2,\cdots,D_{\mathrm{b}}-1\right\}  $. For a product state
having a density matrix $\rho$ given by%
\begin{equation}
\rho=\rho^{\left(  \mathrm{a}\right)  }\otimes\rho^{\left(  \mathrm{b}\right)
}\;,
\end{equation}
the elements of the $D_{\mathrm{a}}D_{\mathrm{b}}\times D_{\mathrm{a}%
}D_{\mathrm{b}}$ Weyl matrix $\mathcal{W}$ (\ref{NLS TS Weyl matrix}) are
given by [see Eq. (8.242) of Ref. \cite{Buks_QMLN}]%
\begin{equation}
\mathcal{W}_{\left(  n^{\prime},n^{\prime\prime\prime}\right)  ,\left(
n^{\prime\prime},n^{\prime\prime\prime\prime}\right)  }=\mathcal{W}%
_{n^{\prime}n^{\prime\prime}}^{\left(  \mathrm{a}\right)  }\mathcal{W}%
_{n^{\prime\prime\prime}n^{\prime\prime\prime\prime}}^{\left(  \mathrm{b}%
\right)  }\;, \label{NLS TS W_n'n'''n''n''''}%
\end{equation}
where $\mathcal{W}^{\left(  \mathrm{a}\right)  }$ and $\mathcal{W}^{\left(
\mathrm{b}\right)  }$ are the Weyl matrices of subsystems a and b, respectively.

Consider a pure state of the composed system given by $\left\vert
\psi\right\rangle \dot{=}\left(  \psi_{1},\psi_{2},\cdots,\psi_{D_{\mathrm{a}%
}D_{\mathrm{b}}}\right)  ^{\mathrm{T}}$. The state vector $\left\vert
\psi\right\rangle $ is represented by a $D_{\mathrm{a}}\times D_{\mathrm{b}}$
state matrix $M$ given by Eq. (\ref{state matrix}) of appendix \ref{AppSMEE}.
For the case where $D_{\mathrm{a}}=D_{\mathrm{b}}\equiv D$, the $D^{2}\times
D^{2}$ Weyl $\mathcal{S}$\ matrix is defined by%
\begin{equation}
\mathcal{S}_{n^{\prime}+n^{\prime\prime}D,n^{\prime\prime\prime}%
+n^{\prime\prime\prime\prime}D}=\frac{1}{D}\operatorname{Tr}\left(  \left(
W_{n^{\prime},n^{\prime\prime}}\otimes W_{n^{\prime\prime\prime}%
,n^{\prime\prime\prime\prime}}\right)  \rho\right)  \;, \label{SNL TS Weyl S}%
\end{equation}
where $n^{\prime},n^{\prime\prime},n^{\prime\prime\prime},n^{\prime
\prime\prime\prime}\in\left\{  0,1,2,\cdots,D-1\right\}  $, and $\rho$ is the
system's density matrix. The PSD matrices $\mathcal{M}=\left(  M^{\dag}M^{{}%
}\right)  \otimes\left(  M^{\dag}M^{{}}\right)  $ and $\mathcal{S}^{\dag
}\mathcal{S}^{{}}$ share the same trace $\operatorname{Tr}\mathcal{M}%
=\operatorname{Tr}\left(  \mathcal{S}^{\dag}\mathcal{S}\right)  =1$, and the
same eigenvalues (i.e. they are unitarily equivalent) [see Eq. (8.245) of Ref.
\cite{Buks_QMLN}]. The parameter $\operatorname{Tr}\left(  \left(
\mathcal{S}^{\dag}\mathcal{S}\right)  ^{2}\right)  $ can be used to quantify
entanglement of a pure state $\left\vert \psi\right\rangle $, for a system
composed of two subsystems, each having Hilbert space dimensionality $D$. For
any fully disentangled state $\operatorname{Tr}\left(  \left(  \mathcal{S}%
^{\dag}\mathcal{S}\right)  ^{2}\right)  =1$, whereas $\operatorname{Tr}\left(
\left(  \mathcal{S}^{\dag}\mathcal{S}\right)  ^{2}\right)  =1/D^{2}$ for any
fully entangled state. For the implementation of disentanglement using the
modified Schr\"{o}dinger equation (\ref{MSE Theta}), the following relation
can be used%
\begin{equation}
\operatorname{Tr}\left(  \left(  \mathcal{S}^{\dag}\mathcal{S}\right)
^{2}\right)  =\left\langle \psi\right\vert T_{2}\left\vert \psi\right\rangle
\;,
\end{equation}
where the operator $T_{2}$\ is given by%
\begin{align}
T_{2}  &  =\frac{1}{D^{4}}\sum_{n_{1},n_{2},\cdots,n_{8}=0}^{D-1}\left(
W_{n_{3},n_{4}}^{\dag}\otimes W_{n_{1},n_{2}}^{\dag}\right)  \rho\nonumber\\
&  \times\left(  W_{n_{3},n_{4}}^{{}}\otimes W_{n_{5},n_{6}}^{{}}\right)
\rho\nonumber\\
&  \times\left(  W_{n_{7},n_{8}}^{\dag}\otimes W_{n_{5},n_{6}}^{\dag}\right)
\rho\nonumber\\
&  \times\left(  W_{n_{7},n_{8}}^{{}}\otimes W_{n_{1},n_{2}}^{{}}\right)
\;,\nonumber\\
&
\end{align}
and $\rho=\left\vert \psi\right\rangle \left\langle
\psi\right\vert $ is the $N^{2}\times N^{2}$ density matrix associated with
the pure state $\left\vert \psi\right\rangle $.

\section{The Schr\"{o}dinger--Langevin equation}

\label{AppSLE}

While the GKSL master equation (\ref{GKSL}) of appendix \ref{AppDamping},
which governs the time evolution of the reduced density matrix $\rho$, is
deterministic, the effect of damping on the time evolution of the state vector
$\left\vert \psi\right\rangle $ can be accounted for by a stochastic equation
of motion, which is known as the Schr\"{o}dinger--Langevin equation, and which
is given by \cite{Jacobs_279,Zhou_272,Liu_024303}%
\begin{equation}
\frac{\mathrm{d}\left\vert \psi\right\rangle }{\mathrm{d}t}=\left[
-i\hbar^{-1}\mathcal{H}+\sum_{l}\left(  \xi_{l}\left(  t\right)
\mathcal{V}_{l}-\frac{1}{2}\mathcal{V}_{l}^{\dag}\mathcal{V}_{l}^{{}}\right)
\right]  \left\vert \psi\right\rangle \;. \label{SLE}%
\end{equation}
The random functions of time $\xi_{l}\left(  t\right)  $, which represent
white noise, have vanishing averaged values $\overline{\xi_{l}\left(
t\right)  }=0$, and correlation functions\ given by%
\begin{equation}
\overline{\xi_{l^{\prime}}^{{}}\left(  t^{\prime}\right)  \xi_{l^{\prime
\prime}}^{\ast}\left(  t^{\prime\prime}\right)  }=\delta_{l^{\prime}%
,l^{\prime\prime}}\delta\left(  t^{\prime}-t^{\prime\prime}\right)  \;,
\label{SLE <xi'*xi''>}%
\end{equation}
where overbar denotes time averaging. For the under study two spin system, the
time independent operators $\mathcal{V}_{l}$ for spin $\mathrm{L}$\ are
$\hbar^{-1}\sqrt{\left(  \hat{n}_{0}^{\left(  \mathrm{L}\right)  }+1\right)
\Gamma_{1}^{\left(  \mathrm{L}\right)  }}S_{\mathrm{L},-}$, $\hbar^{-1}%
\sqrt{\hat{n}_{0}^{\left(  \mathrm{L}\right)  }\Gamma_{1}^{\left(
\mathrm{L}\right)  }}S_{\mathrm{L},+}$ and $\hbar^{-1}\sqrt{2\left(  2\hat
{n}_{0}^{\left(  \mathrm{L}\right)  }+1\right)  \Gamma_{\varphi}^{\left(
\mathrm{L}\right)  }}S_{\mathrm{L},z}$, where $\mathrm{L}\in\left\{
\mathrm{a},\mathrm{b}\right\}  $ [see Eq. (\ref{Lindbladian}) of appendix
\ref{AppDamping}].

\section{Correlation suppression}

\label{AppDO}

Consider a multipartite system composed of three subsystems labeled as 'a',
'b' and 'c'. The Hilbert space of the system $H=H_{\mathrm{a}}\otimes
H_{\mathrm{b}}\otimes H_{\mathrm{c}}$ is a tensor product of subsystem Hilbert
spaces $H_{\mathrm{a}}$, $H_{\mathrm{b}}$ and $H_{\mathrm{c}}$. The
dimensionality of the Hilbert space $H_{\mathrm{L}}$ of subsystem $\mathrm{L}%
$, which is denoted by $D_{\mathrm{L}}$, where $\mathrm{L}\in\left\{
\mathrm{a},\mathrm{b},\mathrm{c}\right\}  $, is assumed to be finite. The PSD
operator $\mathcal{Q}_{\mathrm{ab}}^{\left(  \mathrm{D}\right)  }$ , which is
defined below in Eq. (\ref{Q_12 Tr}), can be used in both the modified
Schr\"{o}dinger equation (\ref{MSE Theta}), and in the modified master
equation (\ref{MME Theta}), to suppress correlation between subsystems a and b.

Let $\left\{  \lambda_{1}^{\left(  \mathrm{L}\right)  },\lambda_{2}^{\left(
\mathrm{L}\right)  },\cdots,\lambda_{D_{\mathrm{L}}^{2}-1}^{\left(
\mathrm{L}\right)  }\right\}  $ be a basis spanning the SU($D_{\mathrm{L}}$)
Lie algebra corresponding to subsystem $\mathrm{L}$, where $\mathrm{L}%
\in\left\{  \mathrm{a},\mathrm{b},\mathrm{c}\right\}  $. In the current study,
both generalized Gell-Mann (see appendix \ref{AppBM}) and Weyl (see appendix
\ref{AppWB}) bases are used for the construction of the decorrelating operator
$\mathcal{Q}^{\left(  \mathrm{D}\right)  }$, which is given by%
\begin{equation}
\mathcal{Q}_{\mathrm{ab}}^{\left(  \mathrm{D}\right)  }=\eta_{\mathrm{ab}%
}\operatorname{Tr}\left(  C^{\mathrm{T}}\left\langle C\right\rangle \right)
\;. \label{Q_12 Tr}%
\end{equation}
The $\left(  a,b\right)  $ entry of the $\left(  D_{\mathrm{a}}^{2}-1\right)
\times\left(  D_{\mathrm{b}}^{2}-1\right)  $ matrix $C$ is the observable
$\mathcal{C}\left(  \lambda_{a}^{\left(  \mathrm{a}\right)  },\lambda
_{b}^{\left(  \mathrm{b}\right)  }\right)  $, and the $\left(  a,b\right)  $
entry of the $\left(  D_{\mathrm{a}}^{2}-1\right)  \times\left(
D_{\mathrm{b}}^{2}-1\right)  $ matrix $\left\langle C\right\rangle $ is its
expectation value $\left\langle \mathcal{C}\left(  \lambda_{a}^{\left(
\mathrm{a}\right)  },\lambda_{b}^{\left(  \mathrm{b}\right)  }\right)
\right\rangle $. For any given observable $O_{\mathrm{a}}^{{}}=O_{\mathrm{a}%
}^{\dag}$ of subsystem a, and a given observable $O_{\mathrm{b}}^{{}%
}=O_{\mathrm{b}}^{\dag}$ of subsystem b, the observable $\mathcal{C}\left(
O_{\mathrm{a}},O_{\mathrm{b}}\right)  $ is defined by $\mathcal{C}\left(
O_{\mathrm{a}},O_{\mathrm{b}}\right)  =O_{\mathrm{a}}\otimes O_{\mathrm{b}%
}\otimes I_{\mathrm{c}}-\left\langle O_{\mathrm{a}}\otimes I_{\mathrm{b}%
}\otimes I_{\mathrm{c}}\right\rangle \left\langle I_{\mathrm{a}}\otimes
O_{\mathrm{b}}\otimes I_{\mathrm{c}}\right\rangle $, where $I_{\mathrm{L}}$ is
the $D_{\mathrm{L}}\times D_{\mathrm{L}}$ identity matrix, and where
$\mathrm{L}\in\left\{  \mathrm{a},\mathrm{b},\mathrm{c}\right\}  $. Note that
the nonnegative expectation value $\tau_{\mathrm{ab}}$, which is given by%
\begin{equation}
\tau_{\mathrm{ab}}=\left\langle \mathcal{Q}_{\mathrm{ab}}^{\left(
\mathrm{D}\right)  }\right\rangle \;, \label{tau_ab}%
\end{equation}
is invariant under any single subsystem unitary transformation. The positive
constant $\eta_{\mathrm{ab}}$ [see Eq. (\ref{Q_12 Tr})] is chosen such that
the expectation value $\tau_{\mathrm{ab}}$ is generally bounded by
$\tau_{\mathrm{ab}}\in\left[  0,1\right]  $. For the two spin 1/2 system under
study $\eta_{\mathrm{ab}}=1/3$.

\bibliographystyle{ieeepes}
\bibliography{acompat,Eyal_Bib}

\end{document}